\documentstyle[preprint,aps]{revtex}
\title{Remarks on Universal  Quantum Computer }
\author{Yu Shi\cite{shi1}}
\address{Cavendish Laboratory, University of Cambridge,
 Cambridge CB3 0HE, United Kingdom}
\begin{document}
\draft
\maketitle
\begin{abstract}

According to Deutsch, a universal  quantum Turing machine (UQTM) is  able to 
perform, in  repeating a  fixed unitary transformation on the total system,
an arbitrary unitary transformation on an arbitrary
data state, by including a program as another part of the input state.
We note that if
such a  UQTM really exists, with the program state dependent on the data
state,   and if the prescribed halting scheme is indeed valid, 
then  there would be  no entanglement between  the halt qubit and
other qubits, as pointed out  by Myers. 
If, however, the program is required to be independent of the data, the  
concerned entanglement appears, and is problematic no matter whether the  
halt qubit is monitored or not. We also note that 
for a  deterministic programmable quantum gate array, as discussed by 
Nielson and Chuang,   if the program
is allowed to  depend on the data state, then its existence
has not been ruled out. 
On the other hand, if UQTM exists, it can be simulated by
repeating the operation of  a fixed gate array.  
However, more importantly, we  observe   that 
it is actually still open  whether Deutsch's  UQTM exists and whether
a crucial concatenation scheme, of which  
the halting scheme is a special case, is valid.

\end{abstract}

 
\section*{1. Introduction}

Quantum computing is under intensive study~\cite{book}.
Most of  the researches follow the approach of
quantum computational network or circuit~\cite{d2}.  
In the present Letter, 
we are  concerned with  quantum Turing
 machine (QTM)~\cite{d1,bernstein} and the so-called universal, 
or programmable, quantum computer.
In classical computation,
Turing machine is employed in stating the Church-Turing thesis,
which is the foundation of the theory of computability: Every computable 
function can
be computed by a   universal  Turing machine. In other words,
all finitely realized computing machines  can be  simulated 
by a {\em single} machine called  ``universal Turing machine''.  
The remarkable  fact 
that one  uses a single computer for various different computational tasks
is a consequence of the Church-Turing thesis.  In the founding paper of quantum
computation~\cite{d1}, Deutsch quantized Turing machine. Furthermore, it
was claimed that there exists a universal QTM (UQTM), which was 
defined to be a QTM  for which there 
always exists a program {\em  as a part of the input state},
which   effects a unitary transformation on an
arbitrary  number of qubits, herein called data qubits,  
arbitrarily close to any desired one. 

The construction of a QTM includes an additional 
halt qubit which signals whether the
computation has completed.  
Recently Myers pointed out that there can
be an entanglement between the halt qubit and the computational 
qubits, hence  monitoring the halt qubit spoils the computation~\cite{myers}. 
Here 
we note that this   situation does not occur 
if UQTM exists and the prescribed
halting scheme is valid, and if the program state
is dependent on the data state.  However, it does occur   if one further
 requires that the program state is independent of the data state,
while  the execution time depends on the data state.  Moreover, the
entanglement between the halt qubit and  the computational qubits is a problem
no matter whether
the halt qubit is monitored or not,  

Deutsch's UQTM is not the only model of universal quantum computer. 
An interesting  and perhaps more practical model is the programmable quantum
gate array discussed by Nielson and Chuang~\cite{nielson}.  Unfortunately 
it was immediately shown  that it does not 
 exist if it is required to operate deterministically. We note that
the disproof is  nullified if the program state is data-dependent. Hence
this model may be more interesting than previously thought.  We also 
note that 
if UQTM exists, one may  use a fixed gate array to simulate each step of
the UQTM, and use  the output of each run of 
the gate array as the input for the next run.

However,  the more important message we
would like to convey is  that in Deutsch's 
 construction of UQTM,
the proof of the existence of 
a program which effects a unitary transformation on the data qubits,
 arbitrarily close to any  desired one, has not actually
been accomplished, and that the concatenation scheme, of 
which the halting scheme is a special case, is also not yet proved to be valid. 

\section*{2. Quantum Turing machine} 

As a generalization  of  classical 
 Turing machine, a QTM
consists of a finite processor consisting of $N$ qubits
$\pmb{n}=\{ n_{i}\}  (i\,=0,\, \cdots, \,N-1)$, and  
an infinite  tape consisting of an infinite sequence of qubits 
$\pmb{m}$ = \{$m_{i}$\}
($i\,=\,\cdots,-1,\,0,\,1,\cdots$), of which
only a finite portion is ever used.
 The currently scanned qubit on the tape, i.e. the position
of the head, is specified by $x$.  Thus the  state of a QTM  
is a unit vector in the Hilbert space 
 spanned by the  basis states
$$|x\rangle|\pmb{n}\rangle|\pmb{m}\rangle,$$
where  $|\pmb{n}\rangle \equiv|n_{0},n_{1},\cdots,n_{N-1}\rangle$, 
$|\pmb{m}\rangle \equiv |\cdots,m_{-1},m_{0},m_{1},\cdots\rangle$.
The dynamics is summarized by a {\em constant} unitary 
operator $U$  whose only nonzero  matrix elements are
$\langle x\pm 1;\pmb{n}';m'_{x},m_{y\neq x}
|U|x;\pmb{n};m_{x},m_{y\neq x}\rangle$. 
{\em  Each different choice of  $U$  defines  a different  QTM}.
To signal whether the computation is completed, another  qubit $n_{h}$ is
added, which   is  initialized to $0$ and is expected to  flip to $1$ 
when the  computation is completed. 
Therefore one may observe 
$n_{h}$  to know whether the computation has been completed.
The evolution of the QTM state can be written as 
\begin{equation}
|\Psi(sT)\rangle = U^{s}|\Psi(0)\rangle,
\end{equation}
where $|\Psi(0)\rangle$ is the initial state, 
$s$ is the number of steps, 
$T$ is the time duration of each  step.

\section*{3. Universal quantum Turing machine}

For a UQTM, we may write its  state  as
 $|{\cal Q}_{x,\pmb{n}}\rangle|n_h\rangle
|{\cal D}\rangle|{\cal P}\rangle|\Sigma\rangle$,
where  $|{\cal Q}_{x,\pmb{n}}\rangle$ is the state of the processor,
including the head position,
$|n_h\rangle$ is the halt qubit,
$|{\cal D}\rangle$ is the state of the data register,
 $|{\cal P} \rangle$ is  the program state.  $|{\cal D}\rangle$ an $
|{\cal P}\rangle$ are both parts of the tape,
$|\Sigma\rangle$ is the rest  part of the tape,  not  affected
during the computation.  Deutsch said that there is a  UQTM, for which  
 there exists a program that effects a unitary transformation, 
arbitrarily close to 
any desired  unitary transformation,  on a finite number of data
qubits~\cite{d1}. In more mathematical details, 
 this can be expressed as the following. 
There exists a special QTM, i.e. a special $U$, 
for an arbitrary unitary transformation ${\cal U}$ and an
arbitrary  accuracy  $\epsilon$, there is always 
a program state 
 $|{\cal P}({\cal D},{\cal U}, \epsilon)\rangle$ and a whole number 
$s({\cal D},{\cal U},\epsilon)$,  such that 
\begin{eqnarray}
U^{s({\cal D},{\cal U},\epsilon)}|{\cal Q}_{x,\pmb{n}}\rangle
|{\cal D}\rangle|{\cal P}({\cal D},{\cal U}, \epsilon)\rangle|\Sigma\rangle
 \nonumber \\
=|{\cal Q}'_{x,\pmb{n}}\rangle
|{\cal D}'\rangle|{\cal P}'({\cal D},{\cal U},\epsilon)\rangle
|\Sigma\rangle, \label{uu}
\end{eqnarray}
where
 $|{\cal D}'\rangle$ is arbitrarily close to $|{\cal U}D\rangle$,
i.e.,
$\left| |{\cal D}'\rangle -|{\cal U}{\cal D}\rangle \right|^2 < \epsilon$.
In the most general case, the program  
${\cal P}$ depends on  the  desired  transformation ${\cal U}$
 and the accuracy $\epsilon$ , as well as
the data state ${\cal D}$.
${\cal P}'$ is the state of the program register after $s$ steps,
${\cal Q}$ and ${\cal Q}'$ are the states of the processor at the initial time
and after $s$ steps, respectively. For the time being, as in Deutsch's
discussion on this issue, we do not consider the  halt qubit, which
will be specifically discussed in Section~5. 

\section*{4. Is there a UQTM?}

Deutsch gave a proof for the above claim.
A key step is the following  inductive  proof of the existence of 
a program which accurately evolves an arbitrary $L$-bit data state 
$|{\cal D}\rangle=|\psi_{1\sim L}\rangle$ to  $|0_{1\sim L}\rangle$, i.e.,
with which $|{\cal D}\rangle$ is evolved to 
$|{\cal D'}\rangle$ which is arbitrarily close to $|0_{1\sim L}\rangle$.
 Here $|0_{i\sim j}\rangle$ denotes
 $|0_{i}\rangle|0_{i+1}\rangle\cdots|0_{j}\rangle$.
Write 
\begin{equation}
|\psi_{1\sim L}\rangle=$$c_{0}|0_{1}\rangle|\psi^{0}_{2\sim L}\rangle
+c_{1}|1_{1}\rangle|\psi^{1}_{2\sim L}\rangle. \label{psil}
\end{equation}
 By inductive hypothesis there exist 
 programs  $\rho_{0}$ and $\rho_{1}$
 which 
 accurately evolve $|\psi^{0}_{2\sim L}\rangle$ 
and $|\psi^{1}_{2\sim L}\rangle$,
 respectively,
  into $|0_{2\sim L}\rangle$. 
 Therefore there exists  a program $\rho$ with the following
effect:  If qubit no. 1 is $0$, execute $\rho_{0}$,
otherwise execute $\rho_{1}$. Thus (\ref{psil}) 
is  converted to
\begin{equation}
\left( c_{0}|0_{1}\rangle+c_{1}|1_{1}\rangle \right) |0_{2\sim L}\rangle,
\label{e4}
\end{equation}
 which can be evolved accurately to $|0_{1 \sim L}\rangle$ by a one-bit
transformation of the qubit no. $1$, using the 
program for one-bit transformations, the existence of which had been claimed
earlier,  using the so-called concatenation, which will be commented on
in the next section.

Let us re-state the above proof in more mathematical details.
The inductive hypothesis is the following. 
There  is a program state $|{\cal P}_0\rangle$ such that 
\begin{eqnarray}
U^{s_0}|{\cal Q}_{x,\pmb{n}}\rangle
|\phi_1\rangle|\psi^0_{2\sim L}\rangle|{\cal P}_0\rangle|\Sigma\rangle
 \nonumber \\
=|{{\cal Q}^0}'_{x,\pmb{n}}\rangle
|\phi_1\rangle|{{\cal D}^0}'\rangle|{\cal P}_0'\rangle
|\Sigma\rangle, \label{uu0}
\end{eqnarray}
while there  is a program state $|{\cal P}_1\rangle$ such that 
\begin{eqnarray}
U^{s_1}|{\cal Q}_{x,\pmb{n}}
|\phi_1\rangle|\psi^1_{2\sim L}\rangle|{\cal P}_1\rangle|\Sigma\rangle
 \nonumber \\
=|{{\cal Q}^1}'_{x,\pmb{n}}\rangle
|\phi_1\rangle|{{\cal D}^1}'\rangle|{\cal P}_1'\rangle
|\Sigma\rangle, \label{uu1}
\end{eqnarray}
where $|{{\cal D}^0}'\rangle$ and $|{{\cal D}^1}'\rangle$ are both 
arbitrarily
close to $|0_{2\sim L}\rangle$, $|\phi_1\rangle$ is a state of the 
qubit no.~1, $s_0$ and $s_1$ are execution times of the two programs,
respectively. 

The existence of 
the  program $\rho$ with the 
effect that if qubit no. 1 is $0$, execute $\rho_{0}$,
otherwise execute $\rho_{1}$ means that there is  a program state 
$|{\cal P}\rangle$ such that $|0_1\rangle|{\cal P}\rangle$ leads to the same
effect on $|0_{2\sim L}\rangle$
as  $|{\cal P}_0\rangle$, while 
$|1_1\rangle|{\cal P}\rangle$ leads to the same
effect on $|0_{2\sim L}\rangle$ as  $|{\cal P}_1\rangle$. 
This can be written as  
\begin{eqnarray}
U^{s_0}|{\cal Q}_{x,\pmb{n}}\rangle
|0_1\rangle|\psi^0_{2\sim L}\rangle|{\cal P}\rangle|\Sigma\rangle
 \nonumber \\
=|{{\cal Q}^0}'_{x,\pmb{n}}\rangle
|0_1\rangle|{{\cal D}^0}'\rangle|{{\cal P}^0}'\rangle
|\Sigma\rangle, \label{uu00}\\
U^{s_1}|{\cal Q}_{x,\pmb{n}}\rangle
|1_1\rangle|\psi^1_{2\sim L}\rangle|{\cal P}\rangle|\Sigma\rangle
 \nonumber \\
=|{{\cal Q}^1}'_{x,\pmb{n}}\rangle
|1_1\rangle|{{\cal D}^1}'\rangle|{{\cal P}^1}'\rangle
|\Sigma\rangle, \label{uu11}
\end{eqnarray}
where $s_0$, $s_1$, $|{{\cal D}^0}'\rangle$ and 
$|{{\cal D}^1}'\rangle$ are the same as those in (\ref{uu0}) and (\ref{uu1}).
Or, in a more relaxed fashion, 
\begin{eqnarray}
U^{s'_0}|{\cal Q}_{x,\pmb{n}}\rangle
|0_1\rangle|\psi^0_{2\sim L}\rangle|{\cal P}\rangle|\Sigma\rangle
 \nonumber \\
=|{{\cal Q}^0}''_{x,\pmb{n}}\rangle
|0_1\rangle|{{\cal D}^0}''\rangle|{{\cal P}^0}''\rangle
|\Sigma\rangle, \label{u00}\\
U^{s'_1}|{\cal Q}_{x,\pmb{n}}\rangle
|1_1\rangle|\psi^1_{2\sim L}\rangle|{\cal P}\rangle|\Sigma\rangle
 \nonumber \\
=|{{\cal Q}^1}''_{x,\pmb{n}}\rangle
|1_1\rangle|{{\cal D}^1}''\rangle|{{\cal P}^1}''\rangle
|\Sigma\rangle, \label{u11}
\end{eqnarray}
where $|{{\cal D}^0}''\rangle$ and $|{{\cal D}^1}''\rangle$ are both 
arbitrarily close to $|0_{2\sim L}\rangle$. 
In general, $s_0 \neq s_1$, $s'_0 \neq s'_1$.

Now we can see two problems. First, there is no reason  why  the existence of
such a  $|{\cal P}\rangle$ automatically  follows the assumption given in
 (\ref{uu0}) and (\ref{uu1}). 
Second,  even if there exists such  
a $|{\cal P}\rangle$, it still does not follow that the decision-making
construction can be realized simultaneously in two branches of a superposed
quantum state, i.e.
\begin{eqnarray}
U^{s}|{\cal Q}_{x,\pmb{n}}\rangle
\left( c_0|0_1\rangle|\psi^0_{2\sim L}\rangle
+c_1|1_1\rangle|\psi^1_{2\sim L}\rangle \right) |{\cal P}\rangle|\Sigma\rangle
 \nonumber \\
=|{\cal Q}_{x,\pmb{n}}'\rangle
\left( c_0|0_1\rangle
+c_1|1_1\rangle \right) |{\cal D}'\rangle |{\cal P}'\rangle
|\Sigma\rangle, \label{up}
\end{eqnarray}
for a certain $s$ and $|{\cal D}'\rangle$ arbitrarily close to 
$|0_{2\sim L}\rangle$.  
What can be known is only that if (\ref{uu00}) and (\ref{uu11}) are valid 
with $s_0$=$s_1$, or if  (\ref{u00}) and (\ref{u11}) are valid 
with $s'_0$=$s'_1$, then
 one can obtain (\ref{up}) by linearly superposing 
(\ref{uu00}) and (\ref{uu11}), or (\ref{u00}) and (\ref{u11}).

In conclusion, in \cite{d1},
although UQTM was defined, its existence was not really proved. 

We do not think that this issue 
 has been solved in a later
work  on   quantum Turing machine~\cite{bernstein}, which 
Myers regards as constrained to special cases. 
Besides,  it seems to us  that in several occasions,  the constructions 
of QTM in \cite{bernstein} rely on a so-called synchronization 
theorem,  which is a property of classical computation. 
But a quantum process
is not a combination  of independent classical processes.  

\section*{5. Concatenation and halting scheme}

In Deutsch's construction, a crucial element is concatenation. The 
concatenation of two programs is a program  whose effect is that of the first
program followed by the second one. It was taken for granted that
if the two programs are valid, their concatenation also exists. 
For example, it was claimed that after obtaining (\ref{e4}), the system
automatically switches to realizing the effect of the  program on
one-bit rotation.  

Suppose there is a program $|{\cal P}_a\rangle$ which effects 
${\cal U}_a$ on a data state $|{\cal D}\rangle$,
and a program $ |{\cal P}_b\rangle$ which effects 
${\cal U}_b$ on  $|{\cal U}_a{\cal D}\rangle$. This means
\begin{eqnarray}
U^{s_a}|{\cal Q}_{x,\pmb{n}}\rangle
|{\cal D}\rangle|{\cal P}_a\rangle|\Sigma\rangle
 \nonumber \\
=|{\cal Q}^a_{x,\pmb{n}}\rangle|{\cal D}_a\rangle|{\cal P}_a'\rangle
|\Sigma\rangle, \label{ua}
\end{eqnarray}
with $|{\cal D}_a\rangle \approx  |{\cal U}_a{\cal D}\rangle$, and 
\begin{eqnarray}
U^{s_b}|{\cal Q}_{x,\pmb{n}}\rangle
|{\cal U}_a{\cal D}\rangle|{\cal P}_b\rangle|\Sigma\rangle
 \nonumber \\
=|{\cal Q}^b_{x,\pmb{n}}\rangle|{\cal D}_b\rangle|{\cal P}_b'\rangle
|\Sigma\rangle, \label{ub}
\end{eqnarray}
with $|{\cal D}_{b}\rangle \approx  |{\cal U}_b{\cal U}_a{\cal D}\rangle$.
But this  does not automatically imply the existence of a program, denoted
as  $ |{\cal P}_{ab}\rangle$, which has the  effect 
\begin{eqnarray}
U^{s_{ab}}|{\cal Q}_{x,\pmb{n}}\rangle
|{\cal D}\rangle|{\cal P}_{ab}\rangle|\Sigma\rangle
 \nonumber \\
=|{\cal Q}^{ab}_{x,\pmb{n}}\rangle|{\cal D}_{ab}\rangle|{\cal P}_{ab}'\rangle
|\Sigma\rangle, \label{uab}
\end{eqnarray}
for a certain $s_{ab}$ and with 
$|{\cal D}_{ab}\rangle\approx  |{\cal U}_b{\cal U}_a {\cal D}\rangle$. 

Therefore the validity of concatenation scheme is not proved. 

The halting scheme is a special case of concatenation scheme. It is based on
the following tacit assumption: if there is a program which effects
${\cal U}$ on the data state $|{\cal D}\rangle$, then there is a program 
which effects $|1_h\rangle\langle 0_h|\otimes {\cal U}$ on 
$|0_h\rangle|{\cal D}\rangle$.
The validity of this assumption is not proved. 

Furthermore, the halting scheme also requires that the halt qubit 
is not flipped
before and after the computation is completed, i.e. it  flips 
only once, when the computation is completed. This 
is a more stringent requirement, the possibility of which is not proved.  

If such a  halting scheme really exists, then in Eq.~(\ref{uu}),
and in  each equation from
 (\ref{uu0}) to (\ref{uab}), one may  add $|0_h\rangle$ on the
left-hand-side and  $|1_h\rangle$ on the
 right-hand-side.

\section*{6. Myers' problem}

Recently, Myers argued
that there can be an entanglement between halt qubit and the 
other qubits, thereby
a measurement on  $n_{h}$  spoils the computation,
as follows\cite{myers}. 
Suppose two  computations, which  start respectively from  basis states 
$|A\rangle|0_{h}\rangle$
and $|B\rangle|0_{h}\rangle$, are  evolved  after $N_{A}$ 
and $N_{B}$ steps   to the desired states  $|A'\rangle|1_{h}\rangle$ and 
$|B'\rangle|1_{h}\rangle$, respectively.
 If $N_{B}\,>\, N_{A}$, then 
for  a computation starting
from $\left( |A\rangle+|B\rangle \right) |0_{h}\rangle$, after  $N$ steps, with
$N_A < N < N_B$, 
the state is  something like
$|A''\rangle|1_{h}\rangle+|B''\rangle|0_{h}\rangle$, because of linearity of
the evolution. In general $|A''\rangle \neq |B''\rangle$. 
Because the computation time is unknown, if one measures $n_{h}$ after 
$N$ steps, with $N_A < N < N_B$, the computation is spoiled since the state
will be reduced to either $|A''\rangle|0_{h}\rangle$ or 
$|B''\rangle|1_{h}\rangle$.
Myers regarded this as a conflict between being universal and being 
fully quantum.

We comment that   the 
combination of  the universality and being fully quantum does  not require  
the whole  quantum computer to evolve from any  superposition;
the arbitrary  
desired  computation ${\cal U}$ is effected only on the data state. 
It is the data state, instead of 
the   state of the total system, that may  start from any  superposition.
In the most general definition of the UQTM, each initial data state
corresponds to a program state, which should contain the control of 
the halt qubit.  When the initial  data state  
is  $|A\rangle+|B\rangle$, for example,  the state of total 
 system starts with 
 $|Q_{x,\pmb{n}}\rangle|0\rangle_h \left( |A\rangle+|B\rangle \right)
|{\cal P}(|A\rangle+|B\rangle,{\cal U},\epsilon)\rangle$, 
rather than 
 $|Q_{x,\pmb{n}}\rangle|0\rangle_h |A\rangle
|{\cal P}(|A\rangle,{\cal U},\epsilon)\rangle+
|Q_{x,\pmb{n}}\rangle|0\rangle_h|B\rangle
|{\cal P}(|B\rangle,{\cal U},\epsilon)\rangle$. 
So  if Deutsch's UQTM really exists and  the halting scheme is really valid,
Myers' problem would  not occur. 

However, Myers' problem does occur if one makes a  further requirement
that the program is only dependent on the desired transformation ${\cal U}$
and the accuracy $\epsilon$, while independent of the initial data state.
Note that this requirement alone does not require that there necessarily 
exists a  program for 
{\em any} desired transformation ${\cal U}$.   In
this case, to compute on a data state
$|A\rangle+|B\rangle$, the system starts with
 $|Q_{x,\pmb{n}}\rangle|0\rangle_h(|A\rangle+|B\rangle)
|{\cal P}({\cal U},\epsilon)\rangle$, which  is equal to 
 $|Q_{x,\pmb{n}}\rangle|0\rangle_h |A\rangle
|{\cal P}({\cal U},\epsilon)\rangle+
|Q_{x,\pmb{n}}\rangle|0\rangle_h|B\rangle
|{\cal P}({\cal U},\epsilon)\rangle$, hence the entanglement 
between the halt qubit and the other qubits appears if the execution time 
is different for $|A\rangle$ and $|B\rangle$. Obviously the problem would be
avoided if the execution time is independent of the initial data 
state. However, this is  
unreasonable, since the number of the data qubits is  arbitrary.  

Our discussions above are from the point of view of whether Myers'
 problem 
appears or not  if Deutsch's original claims are really valid.   
Ozawa considered this issue from  a different angle~\cite{ozawa,ozawa2}.
Based on a   re-formulation of halting scheme, he
argued that Myers' problem does not matter since monitoring 
halt qubit does not change the probability distribution of the output. 
As explained above, we are still
concerned with whether UQTM exists and whether the original 
halting scheme is valid. 
Ozawa's approach is not useful for this question~\cite{m}. 
If, on the other hand, 
Deutsch's constructions of UQTM and the halting  scheme are indeed valid,
then Myers' problem, as  discussed by Ozawa, does not  occur.
If Deutsch's UQTM exists and the halting scheme is valid, and if 
the program is required to be data-independent,  Ozawa's result
still does not solve the problem.
This is because in presence of  the entanglement with the halt qubit,
{\em no matter whether the halt qubit is monitored or not}, 
the computational state is effectively
a mixed one and is not the the pure state designed in
the algorithm.
This is analogous to the situation in a double-slit experiment
where the  which-slit information destroys  the interference~\cite{shi}. 
We also  note that the  halting  scheme does
not necessarily  require that the computational result,  
i.e. $|{\cal D}'\rangle$ in Eq.~(\ref{uu}), remains unchanged under 
the next execution of $U$  on the computer, although such a requirement would
be a nice one.  This is because   in principle  one
could measure the halt qubit at each step,  and stop the unitary 
evolution and  
make measurement immediately when the halt bit flips.  

On the other hand, 
for any QTM, one may regard avoiding the
entanglement between the halt qubit and computational qubits as a
requirement on the dynamics $U$ validating this QTM. 
One can thus see that in general  a same computation 
 with and without a halt qubit correspond to 
different $U$.
For a particular quantum computer,    
it is specifically {\em  devised} in the algorithm, which tells how to
construct the computer,  such   that
there is no need for a halt qubit.  If and only if 
 we consider universal quantum 
computer,    the halting  problem is inevitable. 

\section*{7. 
Universal quantum circuit}

In a quantum circuit, a gate array unitarily transforms a collection of 
qubits from an   input state to an output state. 
In a particular, i.e. non-universal,  quantum circuit, 
as the subject of most quantum computing researches, the desired
unitary transformation  is decomposed into  operations of the   
gates.  
A few  types, or even a single type, of 
gate,   by  repeated use in a repertoire, as prescribed by an algorithm,
can generate the action of any unitary transformation~\cite{book,d2,d3}. 
This is 
another, perhaps currently more popular,
meaning of the word ``universality'', different from what it means 
in our discussions in this paper. 
The total effect of a quantum circuit is simply
\begin{equation}
G|\Psi\rangle_{input} = |\Psi\rangle_{output},
\end{equation}
where $G$ is the unitary transformation, effected by the array of gates.  

Nielson and Chuang discussed a universal quantum circuit~\cite{nielson}, 
which, with a fixed gate array, 
should  perform an arbitrary unitary transformation ${\cal U}$
on a data register $|{\cal D}\rangle$,
by encoding the  program  $|{\cal P}({\cal U})\rangle$
as another part of the input state. The evolution is
\begin{equation}
G \left( |{\cal D}\rangle|{\cal P}({\cal U})\rangle \right) =
|{\cal U}{\cal D}\rangle|{\cal P}'({\cal U})\rangle. \label{nc}
\end{equation}
However,  it was immediately found  that 
such a deterministic 
universal circuit does not exist, since the program states for 
distinct desired  transformations are mutually orthogonal, 
consequently the number of program qubits is infinite for any  finite number
of data qubits .  

Two remarks are in order. First, in the construction and the 
disproof  of Nielson and
Chuang, the program state is assumed to be dependent on the desired 
transformation  ${\cal U}$, but independent of the data state.  
If this assumption is dropped, 
 the proof of the orthogonality of program states for 
distinct desired  transformations is nullified.  Eq.~(\ref{nc})
is changed to 
\begin{equation}
G\left( |{\cal D}\rangle|{\cal P}({\cal U},{\cal D})\rangle\right) =
|{\cal U}{\cal D}\rangle|{\cal P}'({\cal U},{\cal D})\rangle. \label{nc2}
\end{equation}
For $m$ data qubits, 
$2^{m+1}-1$ real numbers are needed to parametrize the data state,
and $2^{2m}$ real numbers are needed 
to parametrize a unitary transformation on 
the data state.
Hence $2^{2m}(2^{m+1}-1)$ real numbers are needed to parametrize the
program state, if the program is dependent on the data state. 
So a universal gate array with data-dependent program
has not been ruled out. One may also allow a finite accuracy $\epsilon$, 
i.e.,
\begin{equation}
G\left(|{\cal D}\rangle|{\cal P}({\cal U},{\cal D},\epsilon\rangle\right) =
|{\cal D}'\rangle|{\cal P}'({\cal U},{\cal D},\epsilon)\rangle, 
\label{nc3}
\end{equation}
with $||{\cal D}'\rangle-|{\cal U}{\cal D}\rangle|^2 < \epsilon$.

The second remark concerns the connection between the universal quantum 
circuit and the UQTM.  If a UQTM exists, one can construct a
fixed gate array such that its effect $G$ corresponds to each step 
$U$ of the UQTM. Then by repeating the operation of  the same fixed
gate array, one can simulate the operation of a UQTM.
The qubits, simulating those of a UQTM, 
are specified as several registers, corresponding respectively to 
the processor state, the halt qubit, 
the data,  the program, and maybe some other  qubits. 
After each  run of the same gate array, the output is taken  
as the input of the next run. The cycle continues until the halt qubit flips,
 and a measurement is then made  on the data state. 
A flow chart is shown in Fig.~1. We may still call this construction a
universal quantum circuit, though it acquires a 
different meaning from  the conventional one.

\section*{8. Summary}

To summarize,  we have discussed several issues regarding quantum Turing
machine, universal quantum Turing machine, and
programmable quantum gate array. 

First, we note that for Deutsch's construction of a
universal quantum Turing machine,  
neither
the claimed  existence of  a program effecting 
 an arbitrary desired  unitary  transformation 
on the data state nor  the validity of the concatenation scheme,
of which  the halting scheme is a special case,  has really been proved. 
Second, we note that  if Deutsch's constructions of 
 universal quantum Turing machine and halting scheme
are really valid, then the    
entanglement between the halt qubit and computational qubits, as observed
by Myers, can be avoided if  the program is allowed to be
 dependent  on the data state. If, however,  the program is
data-independent,  the problem appears no matter whether the halt 
qubit is monitored or not. Finally,
we note that in the programmable quantum gate
array discussed by Nielson and Chuang, if the program state is allowed to 
be dependent on the data state, the disproof of the
deterministic programmable gate array is nullified. Thus  the existence
of a quantum  programmable gate
array with data-dependent program  has not been ruled out.  Furthermore, 
if a  universal quantum Turing machine
 exists, one can simulate it by running  a fixed gate array  
in a cycling way.   In conclusion, it is interesting to make further 
investigations on models of  
universal quantum computer with data-dependent programs. 

\bigskip

I  thank John Myers for interesting discussions. 
Discussions with S. Braunstein, N. Linden, S. Popescu and others are also
acknowledged.  I thank Mike Payne and Peter Littlewood
for hospitality. 

\begin{figure}
\caption{Flow chart of using a fixed gate array
to simulate the UQTM, if the latter exists. 
One run of the fixed gate array, $G$,
corresponds to one step of the UQTM. 
The cycle continues till the halt qubit flips.  $|{\cal D}\rangle$ 
is the input data state,  ${\cal U}$ is an arbitrary desired
unitary transformation on the data state,
 $\epsilon$ is an arbitrary accuracy,
$|{\cal P}({\cal D},{\cal U},\epsilon)\rangle$ is the corresponding
program.  $\left| |{\cal D}'\rangle-{\cal U}|{\cal D}\rangle \right|^2 
< \epsilon$. 
 $|Q(x,\pmb{n})\rangle$   and   $|Q'(x,\pmb{n})\rangle$ are  states
 of the processor, including the  head position, 
 of the simulated UQTM. $|\Sigma\rangle$ represents possible other qubits.} 
\end{figure}

\setlength{\unitlength}{1cm}
\begin{picture}(10,10)\thicklines
\put(1,7){\framebox(9,0.8){$|\Psi\rangle_{input}=|\Psi\rangle_0=|Q(x,\pmb{n})\rangle
|0_h\rangle|{\cal D}\rangle|{\cal P}({\cal D},{\cal U},\epsilon)\rangle
|\Sigma\rangle$}}
\put(2.5,5){\framebox(5,0.9){$i\leftarrow i+1$, $|\Psi\rangle_i=G
|\Psi\rangle_{i+1}$}}
\put(2.5,3){\makebox(5,0.9){Does $n_h=1$?}}
\put(5,3.45){\oval(5,0.9)}
\put(1,1){\framebox(9,0.8){$|\Psi\rangle_{output}=|Q'(x,\pmb{n})
\rangle|1_h\rangle|{\cal D}'\rangle|{\cal P}'({\cal D},{\cal U},\epsilon)
\rangle|\Sigma\rangle$}}
\put(5,7){\vector(0,-1){1}}
\put(5,5){\vector(0,-1){1}}
\put(5,3){\vector(0,-1){1}}
\put(10,6.5){\vector(-1,0){5}}
\put(10,3.5){\line(-1,0){2.5}}
\put(10,3.5){\line(0,1){3}}
\put(5.7,2.5){\makebox(0,0){YES}}
\put(8.7,3.9){\makebox(0,0){NO}}
\end{picture}


\begin{references}
\bibitem{m}  After finishing my work, I heard from 
Myers  that he disagrees with  Ozawa's approach  and   
thinks that one has to also show that the probability distribution without
monitoring the halt flag produces the result one wants. This is 
close to my  comments on  the issue of validity of the halting scheme.
\end{references}
\end{document}